\documentclass[twocolumn,showpacs,preprintnumbers,amsmath,amssymb]{revtex4}

\usepackage{graphicx}
\usepackage{dcolumn}
\usepackage{bm}

\begin{document}

\preprint{APS/123-QED} 

\title{Heavy fermion superconductivity: X-boson treatment}
\author{Lizardo H. C. M. Nunes}
\email{lizardo@if.uff.br}
\author{M. S. Figueira}%
\email{figueira@if.uff.br}
\author{E. V. L. de Mello}
\email{evandro@if.uff.br}
\affiliation{%
Instituto de F\'{\i}sica, C.P.100.093\\
Universidade Federal Fluminense, UFF\\
Av. Litor\^{a}nea s/n, 24210-340 Niter\'oi, Rio de Janeiro, Brasil
}%

\date{\today}
\begin{abstract}
We use the X-boson method to study the heavy fermion 
superconductivity phase employing an extension of
the periodic Anderson model in the
$ U = \infty $ limit with a nearest neighbor attractive interaction between the localized $ f $-electrons. We show that 
higher values of the hybridization parameter
implies in lower values associated to the maximum
of the superconducting critical temperature $ T_{ c } $, 
indicating that the real charge transfer 
between bands tends to destabilizes the Cooper pairing. 
Moreover, we show that the superconductivity
is constrained to the vicinity of a range 
of densities where
the $ f $-band density of states at the Fermi level
$ \rho_{ f } (\mu ) $ have sufficiently high values
and it was 
found both for configurations 
where the system presents intermediate valence and heavy fermion behavior,
as experimentally observed. 
Finally, as the total occupation is raised and for larges 
hybridization parameter $ V $ there is an 
insulator-superconductor transition, 
which is related to
the existence of a hybridization gap that cannot 
be accessed by the slave boson method, because when the chemical potential
lies in or above the gap the system suffers the second order phase transition 
characteristic of that method.
\end{abstract}

\maketitle

\section{Introduction}\label{int}

In the present paper we study  the heavy fermion superconductivity
employing an extension of the periodic Anderson model (PAM) in the
$U=\infty$ limit taking into account the effect of a nearest
neighbor attractive interaction between $f$ electrons
\cite{Matlak87,Romano,portugueses}. The model was primarily
designed to study the heavy fermions systems, but it could
fit in a description of the  high temperature superconductors
compounds (HTSC) as well, if regarded as a simplified extension of
the Emery model \cite{Emery,Dickinson&Doniach}.

Heavy fermion materials present a great variety of ground states:
antiferromagnetic (AF) 
(UAgCu$_{4}$, UCu$_{7}$), 
superconductors
(CeCu$_{2}$Si$_{2}$, UPt$_{3}$),
Fermi liquids (FL)
(CeCu$_{6}$, CeAl$_{ 3 }$) 
and Kondo insulators (KI) 
(Ce$_{3}$B$i_{4}$P$t_{3}$, YbB$_{12}$)
\cite{Steglich1,Fulde88,Hewson93,Aeppli92}. An uniform
high temperature
Curie-like magnetic susceptibility, a common feature to
these compounds, is related to the fact that they are composed of
elements with the incomplete $ f $-shells like $Ce$ and $U$. As
the temperature decreases up to a certain range, the system
presents a temperature independent uniform susceptibility (Pauli
susceptibility) signaling the annihilation or binding of the
magnetic moments of the localized
 $ f $-states, resembling the
single-impurity Kondo problem \cite{Mucio1}. A consistent
description of the overall properties of the heavy fermions is
achieved by the competition between the Kondo effect, dealing with
the annihilation or binding of  the localized magnetic moments,
and the Ruderman Kittel Kasuya Yosida (RKKY) interaction, 
which favors the appearance of a
magnetic ground state; thus, the Hamiltonian that describes the
basic physics of the system may be a regular lattice of $ f $-moments
which interact with an electron gas and with themselves
through these electrons as proposed by the PAM. As concerns the
heavy fermion superconductivity,  it is
an experimental fact that the large specific heat jump at $ T_{c}
$, of the order of the normal phase specific heat, indicates that
pairing takes place among the $ f $-electrons and the
coherent-length must be much shorter than the typical values for
conventional superconductors \cite{Fulde88}. Furthermore, the narrow bandwidth of
the $ f $-electrons suggests a strong on-site Coulomb repulsion
which precludes on-site pairing and may lead to non-$ s $-wave or
unconventional pairing. New results have also revealed the
possibility that magnetic (mainly AF) phase may coexists with the
superconductivity \cite{cond-mat0201040} and that some heavy
fermion compounds display pseudogap behavior at the normal phase
\cite{cond-mat0202251} which is a common feature of the high
critical temperature superconductors \cite{Timusk99}. 

These rich phenomena motivate us to study the
superconducting phase in context of the PAM with the recently
developed X-boson technique, that was applied to the PAM
 in the limit $ U \rightarrow \infty $
\cite{X-boson1,RFranco,FisicaA}. This approach was partially inspired in the
mean-field approximation of Coleman's ``slave boson'' method 
\cite{Coleman84,Coleman87} and
produces satisfactory results, because they are{\ very close to
those obtained by the slave boson in the Kondo limit at low
temperatures, while it recovers those of the chain cumulant
approximation (CHA) \cite{Hewson,Enrique}, which is $\Phi-$
derivable \cite{Baym61,ChainPhi} and is a very good approximation
at high temperature $T$. The unphysical second order phase
transitions that appear in the slave boson approach when
$\mu >> E_{f,j\sigma }=E_{f} $ at low $T$, $ \mu $ being the total
chemical potential and $ E_{f} $ the energy level of the 
$ f $-electrons, and also for
all parameters at intermediate temperatures $T$, are then
eliminated by the X-boson treatment. Coleman \cite{Coleman87} has
observed that these phase transitions are artifacts of the theory,
and the advantage of the present treatment is that those spurious
phase transitions do not occur. Although in the X-boson there is
no relevant spatial dependence in the self-energies of the CHA,
they do retain some local time dependence. In the mean-field slave
boson  the corresponding self-energy vanishes, showing that all
the local time dependence is completely lost in the self-energies
of that method. The spurious phase transition observed by the
slave-boson method are absent from the X-boson description, and
this result indicates that the time dynamics retained in
the CHA is able to suppress those transitions \cite{RFranco}.

In the present paper we adopt the
schematic classification proposed by Varma \cite{Varma85} 
and recently reintroduced by Steglich et al \cite{Steglich1, Steglich2}  
to describe the exemplary Ce-based compounds.
It is given in terms of the dimensionless coupling constant for
the exchange between the local
$ f $ spin and the conduction-electron spins, $g = N_{ F } | J | $, 
where $ N_{ F } $ is the conduction-band density of states at
the Fermi energy and $ J $ is connected to the parameters of the
PAM via a Schrieffer-Wolff transformation \cite{SchriefferWolff},
$J = 2 V^{ 2 } / { E }_{ f } $ when $ U \rightarrow  \infty $. 
Therefore, within the X-boson technique, 
\begin{equation}
g = \rho_{c}(\mu) 2 V^{2} / | \tilde{ E_{ f } } |  \label{g} ,
\end{equation}
where $ \rho_{ c } (\mu ) $ is the conduction density of
states at the chemical potential $ \mu $,
$ V $ is the hybridization parameter and
$ \tilde{ E_{ f } }  $ is the renormalized $ E_{ f } $ 
introduced by the X-boson method, as shall be explained below,
and the behavior of these compounds can be qualitatively
driven through this parameter:
when $g > 1$, the compound under consideration presents an
intermediate valence (IV) behavior while for $ g < 1$ it presents
heavy fermion Kondo regime (HF). There exists a
critical value $ g_{c} $ at which the Kondo and the RKKY
interactions have the same strength and non Fermi-liquid (NFL)
effects have been postulated for systems with $ g = g_{c} $. 
For $g_{c} < g <1 $,
the magnetic local moments do not exist at very low
temperatures and the system presents
a Fermi liquid behavior while for
$g < g_{c}$ we have the local magnetic moment regime.
We observe that we use
the parameter $ g $ to classify the regimes of the PAM in a very
qualitative way. Finally, as concerns to the X-boson chain approach, its
present form only includes hybridization effects to second order
in $ V $, therefore the RKKY effects are not taken into account
and we cannot discuss non
Fermi liquid behavior at the present stage of the work.

The paper is organized as follow: In Section \ref{int} we
introduce the main ideas of heavy fermion superconductivity, in
Section \ref{x-boson} we make a brief revision of the X-boson
method developed earlier \cite{RFranco}, in  Section \ref{super}
we present the model, in Section \ref{Green}  we calculate in a
mean field approximation the Gorkov's anomalous function written
in Zubarev's notation and the gap equation, in Section
\ref{results} we obtain numerically the superconducting phase
diagram for several hybridization parameters and we discuss the
model results; finally in the Section \ref{last} we present the
conclusions and the future directions of the work.

\section{The X-boson method}\label{x-boson}

The periodic Anderson model (PAM) in the limit of infinite Coulomb
repulsion $U\rightarrow\infty$ is given by
\begin{eqnarray}
H & = & \sum_{\mathbf{k},\sigma}\epsilon_{\mathbf{k},\sigma}c_{\mathbf{k},\sigma
}^{\dagger}c_{\mathbf{k},\sigma}+\sum_{j,\sigma}\ E_{f,j\sigma
}X_{j,\sigma\sigma} 
\nonumber\\
& & + \sum_{j,\sigma,\mathbf{k}}\left(
V_{j,\sigma,\mathbf{k}}X_{j,0\sigma
}^{\dagger}c_{\mathbf{k},\sigma}+V_{j,\sigma,\mathbf{k}}^{\ast}c_{\mathbf{k}
,\sigma}^{\dagger}X_{j,0\sigma}\right)  , \label{Eq.3}
\end{eqnarray}
where the first term is the kinetic energy of the
conduction electrons ($c$-electrons), the second term describes
independent localized electrons ($f$-electrons) of energy $ E_{ f } $,
and the last term is the hybridization 
Hamiltonian giving the interaction between
the $c$-electrons and the $f$-electrons.

We employ the Hubbard operators  \cite{Hubbard4} to project the
double occupation state $\left|  \ j,2\right\rangle $ with two
local electrons out from the space of local states at site $j$, as
the $X$ Hubbard operators do not satisfy the usual commutation
relations, the diagrammatic methods based on Wick's theorem are
not applicable, one has to use product rules instead:
\begin{equation}
X_{j,ab}.X_{j,cd}=\delta_{b,c}X_{j,ad}. \label{HO}%
\end{equation}
The identity decomposition in the reduced space of local states at
a site $j$ is
then%
\begin{equation}
X_{j,00}+X_{j,\sigma\sigma}+X_{j,\overline{\sigma}\overline{\sigma}}=I_{j},
\label{Eq.1}%
\end{equation}

\noindent where $\overline{\sigma}=-\sigma$, and the three
$X_{j,aa}$ are the projectors into $\mid j,a\rangle$. Because of
the translational invariance, the occupation numbers
$n_{j,a}=<X_{j,aa}>$ \ satisfy $n_{j,a}$=$n_{a}$ (independent of
j), and from Eq. (\ref{Eq.1}) we obtain the ``completeness''
relation
\begin{equation}
n_{0}+n_{\sigma}+n_{\overline{\sigma}}=1. \label{Eq.2}%
\end{equation}

In Coleman's ``slave boson'' method \cite{Coleman84,Read87}, the
Hubbard X operators are written as a product of ordinary bosons
and fermions:$\ \ X_{j,oo}\rightarrow b_{j}^{+}b_{j}$,
$X_{j,o\sigma}\rightarrow b_{j}^{+}f_{j,\sigma}$, $X_{j,\sigma
o}\rightarrow f_{j,\sigma}^{+}b_{j}$, and a condition, that is
equivalent to Eq. (\ref{Eq.2}), is imposed to avoid states with
two electrons at each site $j$. In the mean field
approximation $b_{i}^{+}\rightarrow<b_{i}^{+}>= \sqrt{z} $ and the method
of Lagrangian multipliers is employed to minimize the free energy
subject to that condition. The problem is then reduced to an
uncorrelated Anderson lattice with renormalized hybridization
$V\rightarrow \sqrt{z}V$ and $f$ level $\varepsilon
_{f}\rightarrow\varepsilon_{f}+\lambda$, and the conservation of
probability in the space of local states is automatically
satisfied because they are described by Fermi operators.

The approximate GF obtained by the cumulant expansion
\cite{FFM,Infinite} do not usually conserve probability (i.e. they
do not satisfy Eq. (\ref{Eq.2})), and the procedure we adopt to
recover this property in the X-boson method is to introduce
\begin{equation}
R\equiv<X_{j,oo}>=<b_{j}^{+}b_{j}>, \label{Eq.5}%
\end{equation}
as variational parameter, and to modify the approximate GF so that
it minimizes an adequate thermodynamic potential while being
forced to satisfy Eq. (\ref{Eq.2}). To this purpose we add to Eq.
(\ref{Eq.3}) the product of each Eq. (\ref{Eq.2}) into a Lagrange
multiplier $\Lambda_{j}$, and employ this new Hamiltonian to
generate the functional that shall be minimized by employing
Lagrange's method. To simplify the calculations we use a constant
hybridization $V$, as well as site independent local energies
${ E }_{f,j,\sigma}={ E }_{f,\sigma}$ and
Lagrange parameters $\Lambda _{j}=\Lambda$. We then obtain a new
Hamiltonian with the same form of Eq. (\ref{Eq.3}):
\begin{eqnarray}
H & = &
\sum_{{\bf {k}},\sigma}\ \epsilon_{{\bf {k}},\sigma}
\ c_{{\bf{k}},\sigma}^{\dagger}c_{{\bf {k}},\sigma} +
\sum_{j,\sigma}\tilde{E}_{f,\sigma}X_{j,\sigma\sigma}
\nonumber \\
& & + V\sum_{j,{\bf {k}},\sigma}
\left(  X_{j,0\sigma}^{\dagger}\ c_{{\bf{k} },\sigma}+
c_{{\bf {k}},\sigma}^{\dagger}\ X_{j,0\sigma}\right)
\nonumber \\
& &+ N_{s}\Lambda(R-1)
\, , \label{Eq.7}
\end{eqnarray}
 but with renormalized localized energies%

\begin{equation}
\tilde{E}_{f,\sigma}=\varepsilon_{f,\sigma}+\Lambda.
\end{equation}

The parameter
\begin{equation}
R=1-\sum_{\sigma}<X_{\sigma\sigma}> \label{Eq. 9}
\end{equation}
is now varied independently to minimize the thermodynamic
potential, choosing $\Lambda$ so that Eq. (\ref{Eq.2}) be
satisfied. While at this stage the electrons in the slave boson
Hamiltonian have lost all the correlations, the Eq. (\ref{Eq.7})
is still in the projected space and it is not necessary to force
the correlations with an extra condition. On the other hand we do
not have an exact solution for this new problem, and we then
consider the most simple approximation obtained within the
cumulant formalism, the Chain approximation (CHA)
\cite{Hewson,Enrique}. The need of minimizing a thermodynamic
potential arises because the completeness relation is not
automatically satisfied for approximate cumulant solutions, and
although the two procedures are formally very similar, they have a
rather different meaning.

\section{The superconductivity model}\label{super}

In order to have a superconducting state, we add to the usual PAM
given by Eq. (\ref{Eq.3}), an effective interaction $H_{W}$ among
the heavy $ f $-electrons,
\begin{equation}
H_{W}=\frac{1}{2} \sum_{<i,j>,\sigma,\sigma'} W_{i,j}
X_{i,\sigma\sigma} X_{j,\sigma'\sigma'}\label{Hj} ,
\end{equation}
where
\begin{equation}
X_{i,\sigma \sigma}=X_{i,0 \sigma}^\dagger X_{i,0 \sigma} \, .
\label{Hj1} 
\end{equation}
The term $H_{W}$ describes an effective attraction between
two neighboring $ f $ sites $ (W_{i,j}<0) $, which is responsible for the
heavy fermion superconductivity in the model. We only consider
the superconductivity arising from $ f $-electron pairing. As the
conduction electrons exist at the Fermi surface the possibility of
formation of $ c $-$ c $ or $ f $-$ c $ Cooper pairs cannot be excluded
\cite{Matlak87}, but we do not consider these
pairs here.
Indeed, there are experimental evidence that
pairing occurs between the heavy $ f $-electrons.
In this framework the correlated
 $ f $-electrons hybridize with the 
conduction band, and interact via a non-retarded
nearest neighbor attraction.
The interaction is proposed
on a phenomenological basis and its microscopic origin
is not investigated. The same approach 
was also used by Romano et. al. \cite{Romano},
Zieli\'{n}ski and Matlak \cite{Matlak87},
and Ara\'{u}jo et. al. \cite{portugueses}. 
Moreover, Tachiki and Maekawa \cite{Tachiki84} studied the
superconductivity in the Periodic Anderson Model
with a small dispersion of the $ f $-band 
in the heavy fermion state
and they have concluded
that pairing between $ f $-electrons are responsible for
superconductivity, rather than 
between the conduction electrons.

It is more interesting to work with the Fourier transform of the
Hubbard operators
\begin{equation}
X_{i,0 \sigma}=\frac{1}{\sqrt{N}}\sum_{l}e^{i\bf {k}.\bf {R_{l}}}
X_{\bf {k},0 \sigma}\label{Fourier1} ,
\end{equation}
\begin{equation}
X_{i,0 \sigma}^\dagger=\frac{1}{\sqrt{N}}\sum_{l'}e^{-i\bf {k}.\bf
{R_{l'}}} X_{\bf {k},0\sigma}^\dagger\label{Fourier2} ,
\end{equation}
where $N$ is the number of the lattice sites. Considering Cooper
pairing only in the singlet channel, the model
defined by Eqs. (\ref{Eq.7}) and  (\ref{Hj}) in the momentum
space can be written as
\begin{eqnarray} \label{model}
H  & = & \sum_{ {\bf k}, \sigma } {\epsilon}_{ {\bf k}, \sigma}
c^{\dagger}_{ {\bf k}, \sigma} c_{ {\bf k}, \sigma} + \sum_{ {\bf
k}, \sigma} \tilde{{E}}_{f} X_{{\bf k},\sigma \sigma }
\nonumber \\
& &+ \sum_{{\bf k}, \sigma } V (X_{\bf
{k},0\sigma}^{\dagger}c_{{\bf k},\sigma} +  c^{\dagger}_{\bf {k},
\sigma}X_{\bf {k}, 0\sigma } )
\nonumber \\
& &+ \sum_{ {\bf k}, {\bf k'} } W_{ {\bf k}, {\bf k'} }
b^{\dagger}_{ {\bf k} } b_{ {\bf k'} } \nonumber \\
& & + N_{s} \Lambda \left( R - 1 \right) \, ,
\end{eqnarray}
where
\begin{equation}\label{bb}
b^{\dagger}_{{\bf k } } = X_{\bf {k},0\sigma}^{\dagger} X_{-\bf
{k},0\overline{\sigma}}^{\dagger}
\end{equation}
and
\begin{equation}\label{interac}
W_{\bf {k},\bf {k'}}=W \sum_{\bf{\delta}} e^{ i(\bf {k}-\bf {k'}).
\vec{ {\delta} } },
\end{equation}
where the summation over $\mathbf{\delta}$ runs over the nearest
neighbors and we consider the hybridization constant
$V=V_{j,\sigma,\mathbf{k}}$.

The superconducting term in 
the right hand side of the Eq. (\ref{model}) is
responsible for the Cooper pair formation. 
The parameter $W$ is negative,
where $ W_{ {\bf k} {\bf k'} } = W \eta_{\bf k}  \eta_{\bf k'}  $
and $ \eta_{\bf k} $ is assigned according to the symmetry of the
superconducting order parameter \cite{portugueses}.

For most of the superconducting materials the charge carriers can
couple in the $ s $, $ p $, $ d $, etc. channels. For heavy
fermions, although, both the crystalline anisotropy and spin-orbit
interaction are important \cite{Yip&Garg}, and the order parameter
should be written in terms of a complete set of basis-function
multiplets for the appropriate symmetries. However, as a first
study of the effect of the hybridization over the superconducting
phase diagram and how the hybridization affects $ T_{ c } $, we
only consider the case of an isotropic $ s $-wave superconducting
gap in our numerical calculations. 

\section{The Green's functions} \label{Green}

Since our model Hamiltonian, Eq. (\ref{model}), does not depend
explicitly on the  time, the corresponding Green's functions GF are
only functions of the difference $ t - t' = \tau  $. In the
aforementioned Zubarev's notation \cite{Zubarev} the GF can be
written as $ G^{c c}_{\sigma }( {\bf k }, \tau ) = \theta( \tau )
\langle c_{ {\bf k}, \sigma}( \tau )    c^{\dagger}_{ {\bf k},
\sigma}(0) \rangle - \theta( -\tau ) \langle c^{\dagger}_{ {\bf
k}, \sigma}( 0 ) c_{ {\bf k}, \sigma}( \tau )    \rangle $ for
this fermionic system, where the operator $ c^{\dagger}_{ {\bf k},
\sigma}( t ) $ is in the Heisenberg representation. Formally,
besides the $ ( {\bf r }, t ) $ representation of the GF, the
model admits the $ ( {\bf r}, \omega ) $ representation defined by the
Fourier transform $ \tilde{G^{ c c } }_{\sigma }( { \bf r }, \omega )
$.
Hence the $ ({ \bf k }, \omega ) $ representation is defined by the discrete Fourier transform
\begin{equation} \label{Gwr}
 \tilde{G^{ c c } }_{\sigma }( { \bf r }, \omega ) =
 \frac{1}{N} \sum_{ \bf k }
 { \mathcal{G } }^{ c c }_{\sigma }( { \bf k }, \omega ) e^{ i { \bf k } . { \bf r } } \, ,
\end{equation}
which could be simply denoted by 
\begin{equation}
 { \mathcal{G } }^{ c c }_{\sigma }( { \bf k }, \omega )
\equiv \ll  c_{ {\bf k}, \sigma} , c^{\dagger}_{ {\bf k}, \sigma}
\gg_{\omega} \, .
\label{EqGcc}
\end{equation}
Analogously, the Gorkov's anomalous function is defined by
\begin{equation} \label{F}
{ \mathcal{ F } }_{ f f, \sigma }^{ \dagger}( { \bf k }, \omega )
\equiv \ll X_{\bf {k},0\sigma}^{\dagger};X_{-\bf
{k},0\overline{\sigma}}^{\dagger} \gg_{ \omega }  \, .
\end{equation}
In a similar way one defines
\begin{equation}
 { \mathcal{ G } }^{ f f }_{\sigma }( { - \bf k }, \omega ) \equiv
 \ll
X_{-\bf {k},0\overline{\sigma}} ;  X_{-\bf
{k},0\overline{\sigma}}^{\dagger} \gg_{ \omega } \, ,
\end{equation}
\begin{equation}
{ \mathcal{ G } }^{ c f }_{\sigma }( -{ \bf k }, \omega ) \equiv \ll
c_{ -{\bf k }, \overline{\sigma}} , X_{-\bf
{k},0\overline{\sigma}}^{\dagger}\gg_{ \omega } \, ,
\end{equation}
and
\begin{equation}
{ \mathcal{ F } }^{ \dagger }_{ c f, \sigma }( { \bf k }, \omega ) \equiv
\ll
 c^{\dagger}_{ {\bf k }, \sigma },
X_{-\bf {k},0\overline{\sigma}}^{\dagger} \gg_{ \omega } \, .
\end{equation}

>From the above definitions, one can show that $ { \mathcal{ F }
}_{ f f, \sigma }^{ \dagger}( { \bf k }, \omega ) $ satisfies the
equation of motion
\begin{equation} \label{moveq}
i \omega { \mathcal{ F } }_{ f f, \sigma }^{ \dagger}( { \bf k }, \omega ) =
\ll Z_{{\bf k },\sigma} ; X^{\dagger}_{-\bf
{k},0\overline{\sigma}} \gg_{\omega} \, ,
\end{equation}
where $ Z_{ {\bf k }, \sigma } = \left[X^{\dagger}_{ \bf
{k},0\sigma}, H \right] $ and $ H $ is given by Eq. (\ref{model}).
We employ the equation of motion method and we obtain a chain of
equations that shall be broken by a mean-field approximation and
further calculations furnish the quasi-particles spectrum.
Therefore, the final
system of equations to be solved is
\begin{eqnarray}
( i \omega_{n} - {\epsilon}_{ {\bf k } } )
{ \mathcal{ G } }^{ c f }_{\sigma } ( -{ \bf k }, \omega_{ n } )
& = &
V { \mathcal{G } }^{ f f }_{\sigma }( { -\bf k }, \omega_{ n } ) \, ,
\label{Gcf}
\\
( i \omega_{n} + {\epsilon}_{ {\bf k } }  )
{ \mathcal{ F } }^{ \dagger }_{ c f, \sigma }( { \bf k }, \omega_{ n } )  &
=  &
- V { \mathcal{ F } }_{  f f, \sigma }^{ \dagger}( { \bf k },
\omega_{ n } )  \, ,
\label{Fcf}
\\
( i \omega_{n} + \tilde{ E_{ f } } )
{ \mathcal{ F } }_{ f f, \sigma }^{ \dagger}( { \bf k }, \omega_{ n } )
& = &
- V D_{ \sigma }  { \mathcal{ F } }^{ \dagger }_{ c f, \sigma }( { \bf k }, \omega_{n } )
      \nonumber \\
& &+
{ \Delta }^{ * }_{ k } D_{ \sigma }
{ \mathcal{G } }^{ f f }_{\sigma }( { -\bf k }, \omega_{ n } )  \, ,
\nonumber \\
\label{Fff}
\\
( i \omega_{n} - \tilde{ E_{ f } } ) { \mathcal{G } }^{ f f }_{\sigma
}( { -\bf k }, \omega_{ n } ) & = & V D_{\overline{\sigma}}
{ \mathcal{ G } }^{ c f }_{\sigma }( -{ \bf k }, \omega_{ n} )
\nonumber \\
& & + \Delta_{k} D_{\overline{\sigma}}{ \mathcal{ F
} }_{ f f, \sigma }^{ \dagger}( { \bf k }, \omega_{ n } )
\nonumber \\
& &+ D_{\overline{\sigma }} \, ,
\label{Gff}
\end{eqnarray}
where $ \Delta_{ k } $ is the superconducting gap defined
by
\begin{equation}
\Delta_{ \bf k} = \sum_{ \bf k k' }
W_{ \bf k k'} \langle b^{\dagger}_{\bf k' }\rangle \,
\label{EqDelta}
\end{equation}
and the quantity $ D_{\sigma } $ is given by
\begin{equation}
D_{ \sigma } =  \langle X_{ 0 0 }
+
X_{ \sigma \sigma }
\rangle \, .
\label{EqDsigma}
\end{equation}

In the X-boson method employed here all the correlations are
included in the quantity $ D_{\sigma } $ and, indeed, making $ D_{
\sigma } = 1 $ the above system of equations with the mapping $ V
\leftrightarrow \sqrt{ z } V $ and $ W \leftrightarrow z^{2} W $
is identical to the slave boson's result given in 
Appendix \ref{slave-boson};
similar to those obtained by Ara\'ujo et.
al. \cite{portugueses}.
Further, in the absence of the hybridization term the
$ f $-band and the conduction band decouples and solving the
remaining system of equations for the $ f $-band one formally
recover the BCS result though with a distinct physical meaning,
here the energy $ E_{ f } $ assumes a constant value and the
strong correlations presented are taken into consideration via the
parameter $ D_{ \sigma } $.


Considering the paramagnetic case 
($ D_{ \sigma } = D_{\overline{\sigma }} $)
and solving Eqs. (\ref{Gcf})-(\ref{Gff}) 
for the anomalous GF one find
\begin{equation} \label{Ffsolve}
{ \mathcal{ F } }_{  f f, \sigma }^{ \dagger}( { \bf k }, \omega )  =
\sum^{ 4 }_{ j =1 } \frac{ B_{ j } }{ i \omega_{ n } - \omega_{ j } } \, ,
\end{equation}
where
\begin{eqnarray}
B_{ 2 } & = &
 \Delta_{ \bf k }
 \frac{
          { { D_{\sigma } } }^2
          \left( {  {\omega_2} }^2 -{ {\epsilon_{ \bf k } } }^2  \right)
          }
         {
            2 \omega_2
           \left( {{\omega_2} }^2 - {{\omega_4}}^2 \right)
           }   \, , \label{B2} \\
B_{ 4 } & = &
\Delta_{ \bf k }
\frac{
         { D_{ \sigma } }^{ 2 }
         \left( { {\omega_4} }^2 -{ { \epsilon_{ \bf k } } }^2  \right)
         }
         {
           2 \omega_4
           \left( {{\omega_4}}^2 - {{\omega_2}}^2  \right)
          } \,  ,\label{B4 }
\end{eqnarray}
with $ B_{1} = - B_{ 2 } $ and $ B_{3} = - B_{ 4 } $.
The four poles representing the
above quasi-particle (QP) excitations given by
\begin{eqnarray}
\omega_{ 2 } & = &
\sqrt{ \frac{ \alpha }{ 2 } -
         \frac{ 1 }{ 2 } \sqrt{ {\alpha}^{ 2 } - 4 \beta }
         }  \, ,\label{omega2} \\
\omega_{ 4 }& = &
\sqrt{ \frac{ \alpha }{ 2 } +
         \frac{ 1 }{ 2 } \sqrt{ {\alpha}^{ 2 } - 4 \beta }
          } \, , \label{omega4}
\end{eqnarray}
$ \omega_{ 1 } = - \omega_{ 2 } $ and $ \omega_{ 3 } = \omega_{ 4 } $, where
\begin{equation}
\alpha=\Delta^{2}_{ \bf k } + 2 V^{2} D_{ \sigma } + { \epsilon_{ \bf k } }^{ 2 } +
{ \tilde{ E_{ F } } }^{2}\, ,\label{alpha}
\end{equation}
and 
\begin{equation}
\beta=V^{ 4 } { D_{ \sigma } }^{2} - 2 V^{2} D_{ \sigma } \epsilon_{ \bf k } \tilde{ E_{ F } } +
{ \Delta_{ \bf k } }^{ 2 }{ \epsilon_{ \bf k } }^{ 2 } + { \epsilon_{ \bf k } }^{ 2 }
{ \tilde{ E_{ F } } }^{ 2 } \, . \label{beta}
\end{equation}

Now we shall treat the hybridization and the superconducting terms
($H_{V}$ and $ H_{W}$ respectively) in the model Hamiltonian given
by Eq. (\ref{model}) as external perturbations and calculate the
thermodynamic potential $ \Omega $ through the $\lambda $
parameter integration \cite{Doniach}. One should minimize the free
energy in order to find the Lagrange multiplier in the former
Hamiltonian. But first we will focus on the contribution of $ H_{
W } $ only, which is the only term responsible for the Cooper pair
formation. Our discussion is analogous to imposing $ V = 0 $ in
our model. Under this  constraint the difference of thermodynamic
potential in the superconductor and normal states 
$ {\Omega}_{ s } - {\Omega}_{ n } $
is
\begin{equation}
 {\Omega}_{ s } - {\Omega}_{ n } =
 \int_{0}^{1} d\lambda \langle H'_{ W }\rangle \, ,
\label{EqdifOmega1}
\end{equation}
where
\begin{equation}
H'_{W} = \sum_{ {\bf k}, {\bf k'} }
W_{ {\bf k}, {\bf k'} }
b^{\dagger}_{ {\bf k} } b_{ {\bf k'} } \, .
\label{EqHJOmega}
\end{equation}
By a variable transformation the above difference defined by Eq.
(\ref{EqdifOmega1}) is rewritten as $ \int_{0}^{ W } (dW' / W')
\langle H'_{ W }\rangle $. In a Hartree-Fock approximation the
average $ \langle H'_{ W }\rangle $ can be ``factorized'' and $
\sum_{ \bf k' } \eta_{\bf k } \eta_{\bf k' } b_{ {\bf k'} } =
\Delta^{*}_{\bf k } / W $ according to Eq. (\ref{EqDelta}). Hence
Eq. (\ref{EqdifOmega1}) becomes
\begin{equation}
{\Omega}_{ s } - {\Omega}_{ n } = \sum_{k} \int_{0}^{W}
\frac{ dW'}{ W'}
\Delta^{*}_{\bf k } \langle b^{\dagger}_{ {\bf k} } \rangle \, .
\label{EqdifOmega2}
\end{equation}
For an isotropic $ s $-wave superconducting gap in the
absence of an external magnetic field
$ \Delta_{\bf k } = \Delta = \Delta^{*} $ and
$ \eta_{ \bf k }  = 1 $. Also notice that
$ W^{- 2} = -d W^{-1} /dW $ and
after another variable transformation we
get that
\begin{equation}
{\Omega}_{ s } - {\Omega}_{ n } =
- \int_{ 0 }^{\Delta} d\Delta' \, {\Delta'}^{2} \,
\frac{ d }{ d\Delta'}\left( \frac{1}{ W } \right) \, ,
\label{EqdifOmega3}
\end{equation}
which is the same functional result given by
Fetter and Walecka \cite{F&W}. Notice that at temperatures
higher than or equal to the superconducting critical temperature
$ T \ge T_{ c } $ the superconducting order parameter
is assumed, as in the BCS theory, to be zero.
Therefore
$ {\Omega}_{ s } - {\Omega}_{ n } = 0 $
at $ T = T_{ c } $
and henceforth we shall
disregard the contribution of $ H_{ W } $
in order to calculate the Lagrange multiplier for
the superconducting phase diagram
and only consider the hybridization term
contribution
of the model Hamiltonian as an external perturbation.
Further, for $ T \ge T_{ c } $ the system is in the normal state and
the GFs yields to the previous
result obtained in the CHA of the PAM, obtained
by taking the infinite sum of all diagrams that 
contains only second order cumulants terms. The 
GFs of the CHA have functional form which are
close to the uncorrelated ones ($ U = 0 $), but they cannot
be reduced to them by any change of scale, except for 
$ D_{\sigma} = 1 $, when we recover the slave-boson
GFs if we put $ V \rightarrow \sqrt{ z } V $.
Nevertheless, notice that although the condition that
forces completeness in the CHA is identical to that employed
in the slave-boson method to force $ n_{ f } \le 1 $, it has a 
rather different origin, being only a consequence of using a reduced
set of diagrams in the perturbative expansion, and the departures
from completeness are usually very moderate. In the formalism 
described here, it is this essential difference 
between the  two methods that eliminates the spurious phase transition appearing in the slave-boson method.
Imposing $ \Delta = 0 $ 
at the system of equations
given by Eqs. (\ref{Gcf})-(\ref{Gff}) one find that
\begin{equation} \label{Gff_Tc}
{ \mathcal{G } }^{ f f }_{\sigma }( { -\bf k }, \omega_{ n } ) \left|_{
T = T_{ c } } \right. = \frac{ A }{ i \omega_{ n } - \omega_{ + } } + \frac{
B }{ i \omega_{ n } - \omega_{ - } } \,
\end{equation}
where the poles $ \omega_{ + } $, $ \omega_{ - } $ are
\begin{equation}
\omega_{ \pm } =
\frac{
         \left(
                 \epsilon_{ \bf k } +
                 \tilde{ E }_{ f }
         \right) \pm
         \sqrt{
                  \left(
                           \epsilon_{ \bf k } -
                          \tilde{ E }_{ f }
                  \right)^{2} +
                  4 V^{2} D_{ \sigma }
                 }
       }{ 2 } \, ,
\end{equation}
and the above coefficients are
\begin{eqnarray}
A =
      D_{ \sigma }
      \frac{
              \epsilon_{\bf k } - \omega_{+}
             }
            { \omega_{+} - \omega_{-} } \, ,
\label{A} \\
B = D_{ \sigma }
 \frac{
           \omega_{-} - \epsilon_{\bf k }
         }
         { \omega_{+} - \omega_{-} } \, .
\label{B}
\end{eqnarray}
Also notice that
$$
\left(i\omega_{ n }-\epsilon_{\bf k}\right) V^{-1}{ \mathcal{ G }
}^{ c f }_{\sigma } ( -{ \bf k }, \omega_{ n } )={ \mathcal{G }
}^{ f f }_{\sigma}( { -\bf k }, \omega_{ n } ) \,  \,
$$
and one get that
\begin{eqnarray}
{ \mathcal{ G } }^{ c f }_{\sigma } ( -{ \bf k }, \omega_{ n } ) & = &
\frac{ V D_{ \sigma }}{ \omega_{ + } - \omega_{ - } }
\nonumber \\
& & \times
\left[ \frac{ 1 }{ i \omega_{ n } - \omega_{ + } } - \frac{ 1 }{ i \omega_{ n
} - \omega_{ - }  } \right] \,  . \label{Gcfsolve}
\end{eqnarray}
Moreover, 
\begin{equation}
{ \mathcal{ G } }^{ c c }_{\sigma } ( { \bf k }, \omega_{ n } )  = 
- \frac{ i \omega_{ n } -   \tilde{ E }_{ f } }
{ 
( i \omega_{ n } - \tilde{{ E }}_{ f }  )
( i \omega_{ n } - \epsilon_{\bf k } )
- V^{ 2 } D_{ \sigma } 
 }
\, .
\label{EqGccsolve}
\end{equation}

Since the thermodynamic potential was already calculated in detail
by one of us and collaborators elsewhere \cite{RFranco} treating
the hybridization Hamiltonian $ H_{ V } $ as an external
perturbation, we shall just briefly outline its solution below.
According to the $ \lambda $ parameter integration
\begin{equation}
\Omega = \Omega_{0} + N_{ s } \Lambda \left( R - 1 \right) +
\int_{0}^{1} d\lambda \langle H_{ V }( \lambda ) \rangle_{ \lambda
} \, , \label{Omega}
\end{equation}
where, regarding Eq. (\ref{Gcfsolve}), the average
$ \langle H_{ V }( \lambda ) \rangle_{ \lambda } $ is
\begin{equation}
\langle H_{ V }( \lambda ) \rangle_{ \lambda }= 4 \lambda V^{ 2 }
D_{ \sigma } \frac{ n_{ F }( \omega_{ + } ) - n_{ F }( \omega_{ - } ) }{ \omega_{
+ } - \omega_{ - } } \, , \label{H_V}
\end{equation}
and $ n_{ F } $ is the Fermi function. At the above Eq.
(\ref{Omega}) the quantity $ \Omega_{0} $ is the thermodynamic
potential associated to the unperturbed part of the Hamiltonian at $
T = T_{ c } $ and $  N_{ s } \Lambda \left( R - 1 \right)  $ was
already introduced to impose the completeness relation
(\ref{Eq.1}). Minimizing $ \Omega $ with respect to $ R $, we get
\begin{equation}
\label{Lambda}
\Lambda = \frac{1}{N_{s}}
             \sum_{ {\bf k } \sigma } V^{ 2 }
                            \frac{ n_{ F }( \omega_{+} ) - n_{ F }( \omega _{ - }) }
                                   {
                                     \sqrt{
                                              \left(
                                                       \epsilon_{ \bf k }
                                                       -
                                                      \tilde{ E_{ f } }
                                              \right)^{ 2 } +
                                     4 V^{2} D_{ \sigma }
                                             }
                                    } \, .
\end{equation}

Finally, the average $ \langle b^{\dagger}_{ {\bf k} } \rangle $
can be found from Eq. (\ref{Ffsolve}) and the self-consistent gap
equation is obtained according to the definition in Eq.
(\ref{EqDelta}). Hence, at the superconducting critical
temperature the superconducting gap for an isotropic $ s $-wave
superconductor satisfies the relation
\begin{eqnarray}
1 & = - W { \beta_{c} }^{ -1 } &
\sum_{ {n}, {\bf k} }
\frac{ D_{\sigma}^{2} }{ \omega_{n}^{2} + \epsilon_{\bf k}^{2} }
\nonumber \\
& & \times
\left[ 
\omega_{n}^{4} +
\omega_{n}^{2}
\left(
2 V^{2} +
\epsilon_{\bf k}^{2}
+
{ \tilde{ E_{ f } } }^{2}
\right) \right.
\nonumber \\
& & + \left.
\left(
V^{2} -
\epsilon_{\bf k}
\tilde{ E_{ f }  }
\right)^{2}
\right]
\, .
\label{EqGapTc}
\end{eqnarray}

Notice that Eq. (\ref{EqGapTc}) is identical to
\begin{equation}\label{gapi}
1  = - \frac{ W }{ \beta_{c} } \sum_{ n, {\bf k }}
\left|
{ \mathcal{G } }^{ f f }_{\sigma }( { -\bf k }, \omega_{ n } )
\right|_{ T = T_{ c } }^{2}
\, .
\end{equation}

>From the numerical point of view it is convenient to perform first
the $\bf k$ summation in the Eq. (\ref{gapi}) and we get that
\begin{equation}\label{gapi1}
1  = -\frac{W D_{\sigma}}{\beta_{c}}\sum_{ n }
\frac{1}{\omega_{n}^{2} + \tilde{E}_{f}^{2}} \frac{1}{N}
\sum_{\bf k} \frac{\omega_{n}^{2} + \epsilon_{\bf k}^{2}}
{(\epsilon_{\bf k} - a_{1})^{2} + b_{1}^{2}} \, ,
\end{equation}
with
\begin{equation}\label{a1}
a_{1}=\frac{V^{2} D_{\sigma} \tilde{E}_{f}} {\omega_{n}^{2} +
\tilde{E}_{f}^{2}} \, ,
\end{equation}
and
\begin{equation}\label{b1}
b_{1}=\frac{[4(\omega_{n}^{2} + \tilde{E}_{f}^{2}) (\omega_{n}^{4}
+ (2 V^{2} D_{\sigma} + \tilde{E}_{f}^{2}))\omega_{n}^{2} + 4
V^{4} D_{\sigma}^{2} \omega_{n}^{2}]^{\frac{1}{2}}}
{2(\omega_{n}^{2} + \tilde{E}_{f}^{2})} \, .
\end{equation}
For simplicity we only consider a constant conduction
density of states in numerical calculations,
\begin{equation} \label{Eqsquarband}
\rho( \epsilon_{\bf k }) =
\left\{
    \begin{array}{l}
     \frac{1}{2D} \; ,
     \mbox{ for } -D \le \epsilon_{\bf k } - \mu \le D\\
      0 \; , \mbox{ otherwise }
    \end{array}  \right. \, ,
\end{equation}
and integrating
Eq. (\ref{gapi1})
over the above square band (\ref{Eqsquarband}),
we get that
\begin{equation}
\label{gapi2}
1  = -\frac{W D_{\sigma}}{\beta_{c}}S_{M} ,
\end{equation}
where
\begin{eqnarray}
S_{M} & = & \sum_{n}\frac{1}
{ (\omega_{n}^{2} + \tilde{E}_{f}^{2})}
\biggl\{1 + \frac{1}{2D} \left(\frac{\omega_{n}^{2} + (a_{1}^{2} -
b_{1}^{2})} {b_{1}}\right)
\nonumber \\
& &\times
\left[\arctan \left(\frac{D-a_{1}}{b_{1}}\right) +
\arctan\left(\frac{D+a_{1}}{b_{1}}\right)\right]
\nonumber \\
& &+
\frac{a_{1}}{2D} \ln \left[\frac{b_{1}^{2} +
(D-a_{1})^{2}}{b_{1}^{2} + (D+a_{1})^{2}}\right]\biggr\} \, .
\label{EqSM}
\end{eqnarray}

The self-consistent solution of the Eqs.  (\ref{Eq. 9}),(\ref{Lambda}), 
(\ref{gapi2}) and (\ref{EqSM}) provide the
superconducting phase diagram.

In the next section we calculate the superconducting 
phase diagram for $ T_{ c } $ as a function of the
total electron number $ N_{ t } = N_{ f } + N_{ c } $,
where these occupation numbers are calculated 
for the Green's functions given by Eqs. 
(\ref{Gff_Tc}), (\ref{Gcfsolve}) and (\ref{EqGccsolve})
considering a constant conduction density of states
as defined in Eq. (\ref{Eqsquarband}). 

The results are
\begin{eqnarray}
{ \mathcal{ G } }^{ f f }_{\sigma } ( \omega )   
& = &
-\frac{ D_{ \sigma } }{ \omega - \tilde{ E }_{ f } }
- \left(\frac{ D_{ \sigma } V } { \omega - \tilde{ E }_{ f } }\right)^{2}
\nonumber \\
& & \times
\log
\left[
\frac
{ 
  \left(\omega - \omega^{1}_{ + } \right) \left( \omega - \omega^{1}_{ - }\right) 
}
{
  \left( \omega - \omega^{ 2 }_{ + } \right) \left( \omega - \omega^{ 2 }_{ - } \right)
}
\right] \, , \label{EqGfffinal} \\
{ \mathcal{ G } }^{ c c }_{\sigma } ( \omega )  
& = &
- \frac{ 1 }{ 2 D }
\log
\left[
\frac
{ 
  \left(\omega - \omega^{1}_{ + } \right) \left( \omega - \omega^{1}_{ - }\right) 
}
{
  \left( \omega - \omega^{ 2 }_{ + } \right) \left( \omega - \omega^{ 2 }_{ - } \right)
}
\right] \, ,
\label{EqGccfinal} \\
{ \mathcal{ G } }^{ f c }_{\sigma } ( \omega )  
& = &
- \frac{ V D_{ \sigma }  }{ \omega - \tilde{ E }_{ f }  }
\log
\left[
\frac
{ 
  \left(\omega - \omega^{1}_{ + } \right) \left( \omega - \omega^{1}_{ - }\right) 
}
{
  \left( \omega - \omega^{ 2 }_{ + } \right) \left( \omega - \omega^{ 2 }_{ - } \right)
}
\right] \, ,
\label{EqGfcfinal} 
\end{eqnarray}
where
\begin{eqnarray}
w^{1}_{ \pm } & = & \frac{1}{2}
\left[ 
 \left( 
 \tilde{E}_{ f } - D_{+} 
 \right) \right. \nonumber \\
 & & \pm
 \left.
 \sqrt{ \left( \tilde{E}_{ f } + D_{+} \right)^{ 2 } + 4 D_{\sigma } V^{ 2 } } 
\right]  
\label{Eqomega1} \, , \\
w^{2}_{\pm } & = & \frac{1}{2}
\left[ 
 \left( 
 \tilde{E}_{ f } + D_{ - } 
 \right) \right. \nonumber \\
 & & \pm
\left.
\sqrt{ \left( \tilde{E}_{ f } - D_{ - } \right)^{ 2 } + 4 D_{\sigma } V^{ 2 } } 
\right]  
\label{Eqomega2} \, , \\
\end{eqnarray}
with $D_{\pm}  = D \pm \mu $. 

Also,
\begin{eqnarray}
N_{ f } & = & \frac{1}{\pi }
\int_{ - \infty}^{\infty} d\omega \, n_{F}( \omega ) \, 
\Im\left[ { \mathcal{ G } }^{ f f  }_{\sigma } ( \omega )  \right]
\, ,
\label{EqNf} \\
N_{ c } & = & \frac{1}{\pi }
\int_{ - \infty}^{\infty} d\omega \, n_{F}( \omega )
\, \Im\left[ { \mathcal{ G } }^{ c c }_{\sigma } ( \omega )  \right] 
\, ,
\label{EqNc}
\end{eqnarray}
where $ n_{ F } $ is the Fermi function.

\section{Results and Discussion} \label{results}

In this section we solve the self-consistent system of
equations  Eqs. (\ref{Eq. 9}), (\ref{Lambda}),  (\ref{gapi2}) 
and (\ref{EqSM}) for $ T_{ c }$.
All the energies are expressed in units of $ D $,
numerical calculation are performed making $ E_{ f } = -0.15  $
and the superconducting interaction parameter is $ W = -0.10 $.

In Fig. \ref{Fig3Phases} we present the superconducting phase
diagram for $ T_{ c } $ as a function of $ N_{t} $,
where each $ N_{ t } $ corresponds
to a distinct sample with a certain stoichiometric composition and
the chemical potential is found self-consistently constrained to
the furnished $ N_{ t } $. The superconducting interaction  $ W $
is kept constant for every set of parameters, since its main
effect is simply to raise the $ T_{ c } $ value. 

\begin{figure}[ht]
\includegraphics[clip,width=0.40\textwidth
,angle=0.]{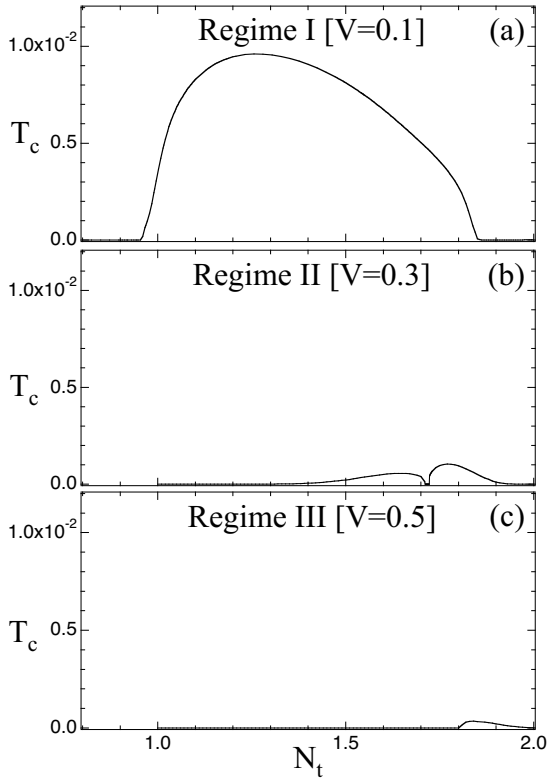}
\caption{The superconducting critical temperature
$T_{ c}$ as a function of the total electron number $N_{ t}$ for
(a) $ V = 0.1 $, (b) $ V= 0.3 $, (c) $ V= 0.5 $ grouped in the three different
regimes. 
$ E_{ f } = -0.15 $ and the superconducting interaction
parameter $ W $ is $ W = -0.10 $, 
all the energies are expressed in units of $ D $.}
\label{Fig3Phases}
\end{figure}

\begin{figure}[ht]
\includegraphics[clip,width=0.34\textwidth
,angle=-90.]{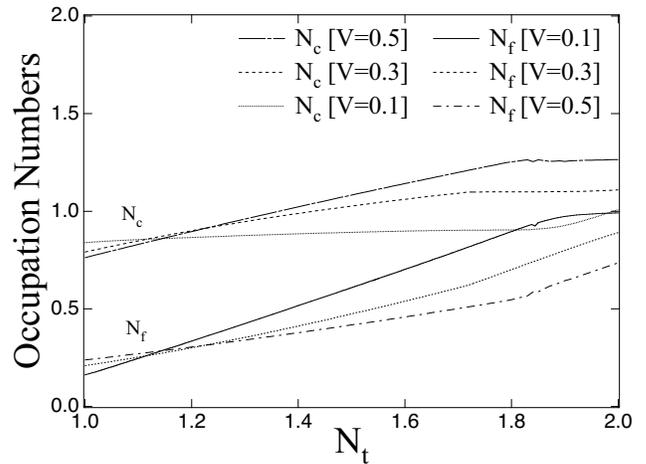}
\caption
{The occupation numbers 
$ N_{ c } $ and  $ N_{ f } $ as a function of $ N_{ t } $.
Other parameters are the same as in Fig. \ref{Fig3Phases}.}
\label{FigNcNfxNt}
\end{figure}

\begin{figure}[htb]
\centerline
{
\includegraphics[width=0.34\textwidth,
angle=90.]
{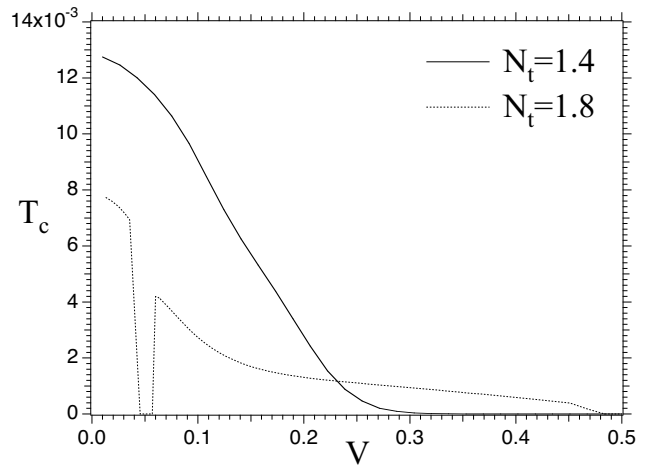}
}
\caption{ The critical temperature $ T_{c } $ as a function of
the hybridization $ V $ for $N_{t} = 1.4 $  and $N_{t} = 1.8 $.
Other parameters are the same as in Fig. \ref{Fig3Phases}.} 
\label{FigTcxV}
\end{figure}

Our  results for the  superconducting phase diagram 
show that the existence of the superconductivity 
is constrained to the vicinity of a 
range of occupations where the values of the 
$ f $-band densities of states 
$ \rho_{f}( \mu ) $ at $ \omega = \mu $ are sufficiently high.
Moreover, high values of $ \rho_{f}( \mu )$ usually 
implicate higher values of $ T_{c} $, what suggests that
the Kondo behavior of the system favors superconductivity
in the X-boson approach. Nevertheless, the 
reduction of the hybridization parameter $ V $ causes a general increasing on the maximum of
the superconducting critical temperature $ T_{ c } $, this is due
to the diminution of the charge fluctuations between the
conduction and the $ f $ electron states.
Indeed, Fig. \ref{FigNcNfxNt} shows the occupation numbers
$ N_{ c } $ and  $ N_{ f } $ as a function of $ N_{ t } $, 
for a given total electron number 
as $ V $ increases, part of the electrons that would be  
in the localized $ f $-band, and therefore responsible for the Cooper pair formation, should be hybridized with the conduction band, showing that the charge fluctuation between the conduction and the 
$ f $-electron states makes the Cooper pairs unstable. Similar results
were also obtained by Romano et al. \cite{Romano} and 
Sarasua and Continentino \cite{LGSarasua}. 
Indeed, Fig. \ref{FigTcxV} shows the behavior 
of $ T_{c } $ as a function of
the hybridization $ V $ in the low and high
occupation regimes. As $ V $ increases, the critical temperatures
diminishes and at very small values of the hybridization,
the critical temperatures tend to a finite value;
this behavior is similar to that obtained by
Ara\'ujo et. al. \cite{portugueses}  
using the slave boson method.
Note that for $ N_{ t }  =  1.8 $ the superconductivity 
phase is suppressed for the interval $ V \sim [0.03, 0.06 ] $, 
which is related to
the appearance of a hybridization gap in the
$ \rho_{ f }(\omega )  $.    

\begin{figure*}[ht]
\includegraphics[clip,width=0.50\textwidth
,angle=-90.]{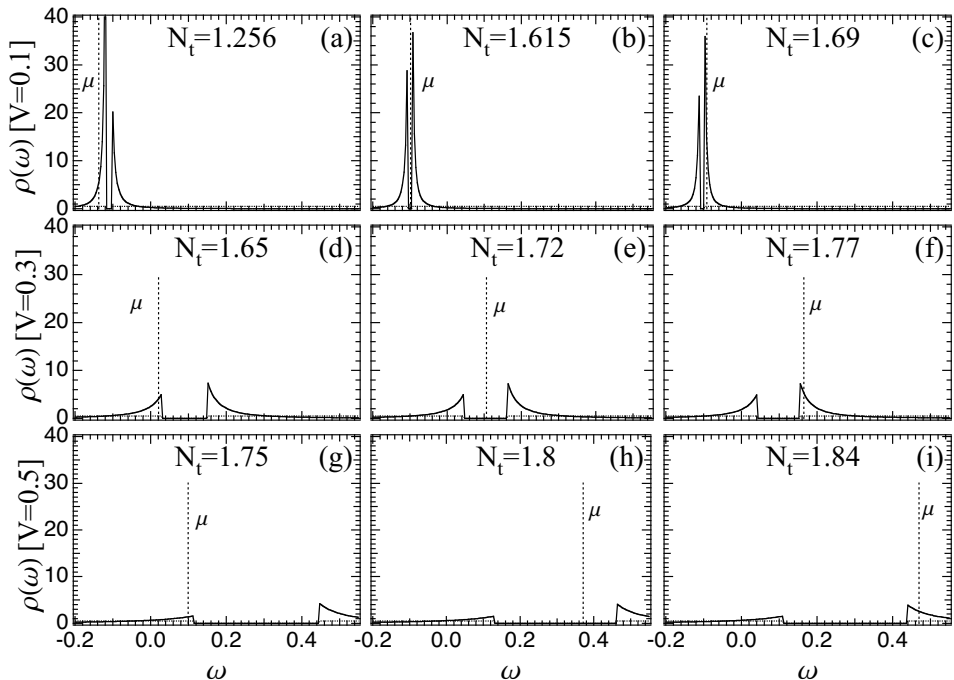}
\caption{Plots of the $ f $-density of states 
$ \rho_{ f  }(\omega) $ (filled line) and the
$ c $-band density of sates $ \rho_{c }(\omega) $ (dashed lines) 
for (a)-(c) $ V=0.1 $,  
(d)-(f) $ V=0.3 $, and (g)-(h) $ V=0.5 $ 
compared to 
the values of the chemical potential 
$ \mu $ (vertical short dashed line).
Other parameters are the same as in Fig. \ref{Fig3Phases}.}
\label{FigDOS}
\end{figure*}

Regarding the superconducting phase diagram, 
at low total occupation $ N_{t} $, the 
$ f $ level occupancy is also small and,  
since Cooper pairing occurs only between the $ f $ electrons,
$ T_{ c} $ is null. 
As the values for $ N_{ t } $ are raised, $ T_{ c } $ increases as
$ \rho_{f}( \mu ) $ is increased and presents its
maximum in the vicinity of the Kondo resonance.
In the large $ N_{ t } $ regime 
the $ f $ level is almost fully occupied 
with one electron per state and the 
$ \rho_{f}( \mu ) $ decreases causing the suppression of
superconductivity, but it should be noted that
$ T_{ c } $ vanishes before $ N_{ f } \rightarrow 1 $.
Similar results were also obtained by Ara\'ujo et al \cite{portugueses},
except that for them the superconducting critical temperature
is constrained to $ T_{ c } \le T_{ K } $, where the Kondo temperature 
$ T_{ K } $ is defined as the $ T $ that makes the slave-boson parameter
$ z  $ vanish. Indeed, in the slave-boson method by increasing the temperature
$ T $ or the chemical potential $ \mu $ the value 
$ \tilde{ V } \equiv \sqrt{ z } V  = 0 $ is presently reached, 
leading to $ N_{ f } \rightarrow 1 $ and the 
decouple of the two types of electrons, what can be interpreted
as a change of phase related to a symmetry breaking of the 
mean-field Hamiltonian. These unphysical second-order phase transitions
that appear in the slave-boson approach are artifacts of the theory, 
as already observed by Coleman \cite{Coleman85},
while the present X-boson treatment prevents those 
spurious phase transitions. 
Therefore, we were motivated to
study the superconducting phase diagram systematically  
by varying the hybridization parameter $ V $ and our results 
provide three distinct regimes for the superconducting phase diagram
as can be seen in Fig. \ref{Fig3Phases}: 
(a) for small values of $ V $ the
superconductivity shows up in a broad interval of occupations and
the function $ T_{c}( N_{t}) $ have a single maximum for our
numerical data; (b) as $ V $ is increased, $ V \sim 0.2-0.4 $, the range of occupations
where there is superconductivity is narrowed and shifted to
higher values. The function $ T_{c}( N_{t}) $ presents two local maxima and
between them the superconductivity is suppressed, which is related to
the appearance of a hybridization gap in the $ \rho_{ f }(\omega )  $
and the system becomes an
insulator in this region; 
(c) as $ V $ increases even more, $ V \sim 0.4 - 0.5 $,
the superconductivity shows 
up in a smaller range of occupation and 
the superconducting region related to the first local maximum
is suppressed. As the total occupation is raised, 
the system crosses an insulating region until it becomes superconducting. 
For larger values of the hybridization parameter
($ V > 0.5 $) the superconductivity is completely suppressed.

\begin{figure}[ht]
\includegraphics[clip,width=0.35\textwidth
,angle=0.]{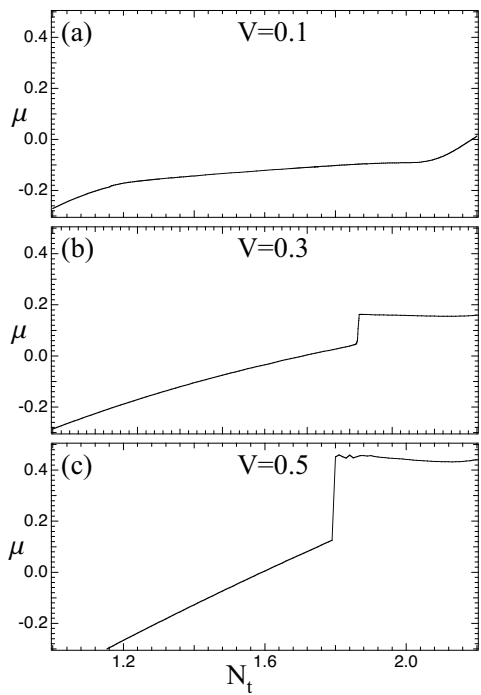}
\caption{The chemical potential
$ \mu $ as a function of $ N_{ t } $.
Other parameters are the same as in Fig. \ref{Fig3Phases}.} 
\label{FigMuxNt}
\end{figure}

According to our data for the
superconducting phase diagram in the first regime,
the function $ T_{c}( N_{t}) $ has
a single maximum, as can be seen in Fig. \ref{Fig3Phases}.a.  
For this low $ V $ regime the $ f $-band density of
states is located in a narrow region of the bandwidth, since 
the  electrons are more localized, the two
peaks of the density of states are sharp and the distance between
them is small, about $ 0.02D $, as can be seen in 
Figs. \ref{FigDOS}.a-\ref{FigDOS}.c.
Notice that in this small interval
defined by the distance between peaks, the densities of states for
both bands are zero and
when the chemical potential lies in this hybridization gap
the system presents an insulating behavior unless
the conducting electrons have enough kinetic energy to
tunnel through the gap. 
In this low hybridization regime the superconducting 
critical temperature is larger than the distance between
the chemical potential $ \mu $ and the border band peaks, 
and since the average kinetic energy of the 
conducting electrons should
be proportional to $ T_{ c } $ for a given 
occupation in the phase diagram, 
the conducting electrons have enough energy to
tunnel through the hybridization gap and the 
system does not present an insulating 
behavior. Notice that $T_{ c }$ is larger for 
$ V = 0.1 $ because the existence of
the superconductivity phase is related to a high value for
the $ f $-band density of states at the chemical potential.

The phase diagram for the second regime
is presented in Fig. \ref{Fig3Phases}.b. 
The critical temperature  $ T_{ c } $ is non-null
only when the $ f $-band density of states at the chemical
potential is sufficiently high, what also indicates the range
of occupations where there is superconductivity.
Notice the two local maxima for $ T_{ c} $ and the
suppression of superconductivity between them 
when $\mu $ crosses the hybridization gap. 
In this region the Matsubara's summation
defined by Eq. (\ref{EqSM}) is close to zero and, hence, there
is no non-null solution for $ T_{ c} $.
Indeed, in Figs. \ref{FigDOS}.d - \ref{FigDOS}.f we
present the results for the chemical potential localization,
the $ f $-band density of states and the conducting electrons
density of states $ \rho_{ c } (\omega ) $ corresponding to
different values of $N_{t}$.
Moreover, Fig. \ref{FigMuxNt} presents the behavior of
the chemical potential as a function of $ N_{ t } $
for the three regimes found. Notice that
there is an abrupt variation of $ \mu $ when it crosses
the hybridization gap, the slope is larger as
the distance between peaks of the DOS
is larger. Indeed, Fig. \ref{FigDOS} shows the appearance
of a hybridization gap when the chemical potential lies in a
region between bands where superconductivity is suppressed
because the system becomes an insulator. 
An analogous transition was found by Romano \cite{Romano}
studying the same model taking into account 
$ s $, $ p $, $ d $ anisotropic pairing and for $ U = 1$,
in units of $ D $
for the value of $ N_{ t } $ close to the half filling.
Indeed, notice that when the  values of the hybridization parameter
are raised the two peaks of the density of states are lowered
and the distance between peaks is broadened, evidencing two main factors responsible for the appearance of 
an insulator region under the constraints of the
furnished parameters in the X-boson approach:
on one hand, when
$ V $ increases, the $ f $-band density takes up smaller values,
what causes the overall superconducting 
critical temperature to diminish, since $ T_{ c} $
is non-null only when the value of $\rho_{f}( \mu ) $
is sufficiently large, indicating that the 
average kinetic energy of the fermions in the system is decreased; on
the other hand, since the distance between peaks 
are broadened, the demand of kinetic energy required
by the conducting electrons to tunnel through
the hybridization gap becomes larger. Therefore,
as the total occupation is increased the
chemical potential roughly also increases, 
as seen in Fig. \ref{FigDOS}, and when it crosses
the hybridization gap, the conducting electrons
do not have the required energy to tunnel
through the gap and the system becomes an insulator.
Again, this result would
not be obtained by the mean-field slave boson method \cite{RFranco},
since the approach breaks down in the Kondo region when 
$ N_{ f } \rightarrow 1 $ and the upper band cannot be reached.

For the third regime, superconductivity
is constrained to a small range of occupation, as can be seen
in Fig. \ref{Fig3Phases}.c.
This regime is characterized
by high values of the 
hybridization parameter and
the superconducting region, related to the first local maximum
of $ T_{ c } $, is suppressed. Therefore,
as the total occupation is raised, 
or charge carries are added to the 
$ f $-band, the system suffers a
insulator-superconductor transition.
Again, our results
for the superconducting phase diagram can be inferred from the
behavior of the chemical potential compared to several plots of
the the $ f $-band density of states.
Indeed, as can be seen in 
Figs. \ref{FigDOS}.g-\ref{FigDOS}.i,
the $ f $-band DOS is asymmetric and the 
first peak, which is related to the first
local maximum of $ T_{ c } $, is smaller;
hence, for this range of occupations
the values of $\rho_{f}( \mu ) $
are not sufficiently large and the superconductivity
is suppressed.

\begin{figure}[htb]
\centerline
{
\includegraphics[width=0.32\textwidth,
angle=90.]
{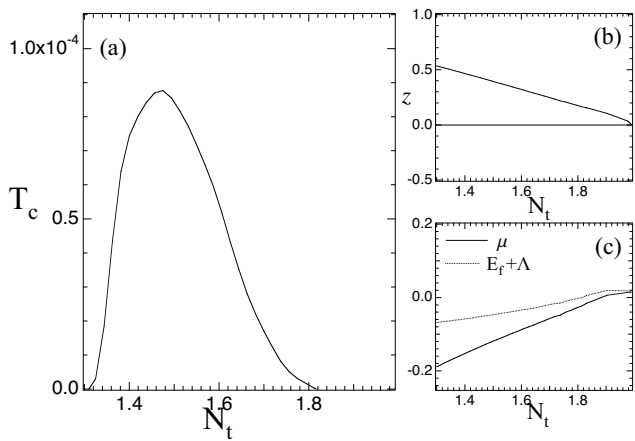}
}
\caption{ (a) Slave boson's result for 
$ T_{ c} $ as a function of $ N_{ t} $.
(b) $ z  $
as a function of $ N_{ t} $. 
(c) The chemical potential $\mu $ and 
the renormalized energy
$ {E}_{f} + \Lambda $ 
as a function of $ N_{ t} $.
$ V = 0.2 D $, $ W = -0.5 D $,
and other parameters are the same as in Fig. \ref{Fig3Phases}. } 
\label{FigSlaveTcxNt}
\end{figure}

It is interesting to compare the results
obtained by the X-boson approach
with those obtained by the slave boson approach 
(see Appendix \ref{slave-boson}):
first note that while the effective superconducting
interaction $ W $ is mapped according to 
$  W \rightarrow D^{2}_{\sigma } W $ for
the X-boson method, the effective $ W $ for the
slave boson method is 
$ W  \rightarrow z^ {2} W $, 
see Appendix \ref{slave-boson}.
Note that $ D_{\sigma} $ is a bounded quantity
varying from 1 to 0.5 as the  
$ f $-band occupation raises,
while $ z \rightarrow 0 $ as
$ N_{ f } \rightarrow 1 $
and the main effect of the $ z^2 W $
interaction is to cause 
a general reduction of  
$ T_c $ for the occupations 
where the superconductivity exists.
Hence, only for a large value of
the $ W $ parameter we could find
a value for $ T_c $ in the slave boson
method which is comparable to those 
obtained by the X-boson approach,
what means that the slave boson
method requires greater values
of $ W $ to produce
the same critical temperatures
reached by the X-boson method, see \cite{portugueses}.
Furthermore, Fig. \ref{FigSlaveTcxNt}.a
presents the superconducting phase
diagram for $ T_{ c } $ as a function of $ N_{t} $ for
$ V = 0.2 $ and $ W = -0.5 $. Note that as
$ N_{t} $ increases, the chemical potential also increases,
but $ z $, which is essentially the expectation value 
of the $ f $-holes, tends to zero
(Fig. \ref{FigSlaveTcxNt}.b)
causing $ T_c $ to be  null
before $ \mu > \tilde{ E }_{f} $
(Fig. \ref{FigSlaveTcxNt}.c).

\begin{figure}[ht]
\includegraphics[clip,width=0.37\textwidth, 
,angle=0.]{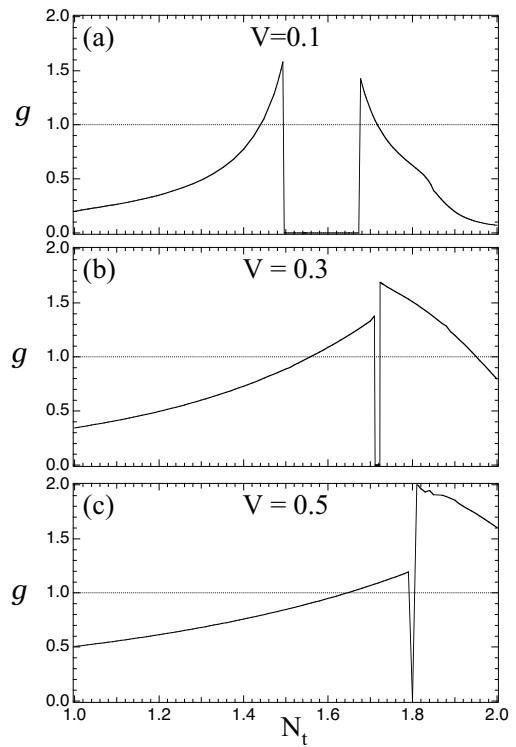}
\caption{Parameter $ g $, as defined by Eq. \ref{g}, as a function of
$ N_{ t } $ for (a) $ V=0.1 $, (b) $ V=0.3 $, and (c) $ V=0.5 $. 
Other parameters are the same as in Fig. \ref{Fig3Phases}.} 
\label{Figg}
\end{figure}

Finally, in Fig. \ref{Figg} we show
the parameter $ g $ defined by the Eq. \ref{g}
as a function of $ N_{ t } $ for
the three regimes. 
As can be seen in Fig. \ref{Figg}, 
according to the criteria
established in \cite{Steglich1,Steglich2}, 
as $ N_{ t} $ increases 
we recover the three
characteristic regimes of the PAM:
Kondo, IV and 
magnetic; the same cannot be found for the
slave-boson treatment, since it breaks down
when $ N_{ f } \rightarrow 1 $.
Also notice that the criteria is consistent to the picture
that as $ \mu $ approaches to $ E_{ f } $
within an interval of width 
$ \Delta_{A } \propto  V^{ 2 } $,
there exists the possibility of charge fluctuations
between the bands.
For $ V = 0.5 $, the superconductivity only
exists for a range of $ N_{ t } $ where the 
system presents a IV behavior ($ g > 1 $). 
As the value of the hybridization is lowered, 
we prevent the possibility of charge fluctuation
between bands for broader ranges of $ N_{ t } $ 
and the superconductivity arises even in the regions where 
the system presents Kondo or magnetic behaviors.
Since high values of $ \rho_{f}( \mu ) $ usually 
imply in higher values of $ T_{c} $, the X-boson approach
favors the superconductivity around the Kondo resonance.
Nevertheless, as $ \mu $ crosses the hybridization gap,
when the chemical potential lies in the region 
between the peaks of the density of states,
the quantity $ \rho_{ c } ( \mu ) \rightarrow 0 $,
and one gets that $ g =0 $, since it is proportional 
to the $ c $-band density of states at the chemical potential
and nothing can be
inferred about the magnetic order the system.
Moreover, as $ N_{ f } \rightarrow 1 $, the parameter
$ g $ decreases. 
Indeed, since
$ T_{RKKY} \sim g^{2} $ and
$ T_{ K } \propto e^{-1/g} $ \cite{Steglich1,Steglich2},
when $ g \ll 1 $ one should expect that
the RKKY interaction, which scales 
to $ k_{B} T_{RKKY} $, to be much larger
than $ k_{B} T_{K} $, suggesting that the magnetic
ordering at $ T_{RKKY} $ pre-empts
Kondo singlet formation, and the system
might go from a FL paramagnetic state
to an RKKY magnetic phase with
well-localized moments according to the 
established criteria.

\section{Conclusions}\label{last}

In conclusion, in this paper we studied within the
X-boson approach 
the paramagnetic case of
the PAM with a phenomenological superconducting term and 
we obtained the phase diagram of the model.
Although high values of $ \rho_{f}( \mu ) $ usually 
imply higher values of $ T_{c} $, what indicates that
the X-boson approach
favors superconductivity around the Kondo resonance,
superconductivity was found both for
configurations where the system presented IV and HF
behavior. This behavior
is in agreement to experimental results,
since superconductivity was found both
in heavy fermion materials as well as
in intermediate valence compounds \cite{IVSC1,IVSC2}.
Moreover, we show that higher values of the
hybridization parameter implies in lower values associated to the
maximum of the superconducting critical temperature, indicating
that the real charge exchange between bands tends to destabilizes
the Cooper pairing, in agreement to the previous results obtained
by the slave-boson approach \cite{portugueses} and by perturbative
approach \cite{Romano}. 
Furthermore, as the total occupation is
raised and for larger $ V $, the system presents 
a superconductor-insulator transition,
which is related to the appearance of a hybridization gap that 
cannot be obtained by the slave-boson method, which breaks down in
this region.
These results could be experimentally tested 
since the hybridization coupling $ V $ can be increased by
an applied external pressure.

Also, the detailed discussion of the magnetic
solutions of the model
within the X-boson approach, the
possibility of coexistence between superconductivity
and magnetic order and even how the 
different symmetries of the superconducting
order parameter can
alter the results obtained in the present paper 
will be subject of investigation in 
future works.

\begin{acknowledgments}
We acknowledge Profs.  M. E. Foglio (Unicamp) and M. A. Continentino (UFF) for helpful
discussions and the financial support from the Rio de Janeiro
State Research Foundation (FAPERJ) and National Research Council
(CNPq).
\end{acknowledgments}

\appendix
\section{Slave boson approach} \label{slave-boson}

We apply the Coleman's ``slave boson'' method \cite{Coleman84}
described in Section \ref{x-boson}
to the Hubbard operators in the original Hamiltonian given in
Eq. \ref{Eq.3}
plus the term 
$\sum_{ {\bf k}, {\bf k'} } 
W_{ {\bf k}, {\bf k'} }
b^{\dagger}_{ {\bf k} } b_{ {\bf k'} }$,
which is responsible for the Cooper pair formation in the singlet state.
Here we only assume  a constant
hybridization $ V $, as well as site independent local energies
${ E }_{f,j,\sigma}={ E }_{f,\sigma}$.
Also, we are constrained to the subspace where the identity
$ I = b^{\dagger } b  + \sum_{ \sigma } f^{ \dagger }_{\sigma} f_{ \sigma } $
is preserved. 
Therefore, the operator 
$ Q \equiv  \Lambda 
\left(
z  +
\sum_{ {\bf k} \sigma } f^{ \dagger }_{\bf k, \sigma } f_{ {\bf k} \sigma } 
 - 1 \right)
$
is added to the mean-field Hamiltonian
rewritten in the slave-boson's representation
and the parameters $ z $ and $ \Lambda $
are determined by minimization of the free
energy.
In terms of this new representation the model Hamiltonian becomes
\begin{eqnarray}
H  & = & 
\sum_{ {\bf k}, \sigma } {\epsilon}_{ {\bf k}, \sigma}
c^{\dagger}_{ {\bf k}, \sigma} c_{ {\bf k}, \sigma} 
+ 
\sum_{ {\bf k}, \sigma} 
\tilde{{E}}_{f} f^{\dagger}_{ {\bf k}, \sigma }f_{ {\bf k}, \sigma }
\nonumber \\
& &
+ \sum_{{\bf k}, \sigma } \sqrt{z}V 
( 
c^{\dagger}_{\bf {k}, \sigma} f_{ {\bf k }, \sigma }
+ 
c_{ {\bf k},\sigma} f^{\dagger}_{ {\bf k }, \sigma }
)
\nonumber \\
& &
+ \sum_{ {\bf k}, {\bf k'} } z^{2} W_{ {\bf k}, {\bf k'} }
b^{\dagger}_{ {\bf k} } b_{ {\bf k'} } \nonumber \\
& &
+ \Lambda \left( z - 1 \right) \, ,
\label{EqSlaveModel}
\end{eqnarray}
where the localized  $ f $-energy is renormalized by
$ \tilde{{E}}_{f} = {E}_{f} + \Lambda $,
and the operator $ b^{\dagger}_{ {\bf k} } $  
is given by
$ b^{\dagger}_{ {\bf k} } = 
f^{\dagger}_{ {\bf k }, \uparrow } 
f^{\dagger}_{ -{\bf k }, \downarrow } $. 

Finally, in a mean-field approximation we find
\begin{eqnarray}
( i \omega_{n} - {\epsilon}_{ {\bf k } } )
{ \mathcal{ G } }^{ c f }_{\sigma } ( -{ \bf k }, \omega_{ n } )
& = &
V \sqrt{z} { \mathcal{G } }^{ f f }_{\sigma }( { -\bf k }, \omega_{ n } ) \, ,
\label{SlaveGcf}
\\
( i \omega_{n} + {\epsilon}_{ {\bf k } }  )
{ \mathcal{ F } }^{ \dagger }_{ c f, \sigma }( { \bf k }, \omega_{ n } )  &
=  &
- V \sqrt{z} { \mathcal{ F } }_{  f f, \sigma }^{ \dagger}( { \bf k },
\omega_{ n } )  \, ,
\label{SlaveFcf}
\\
( i \omega_{n} + \tilde{ E_{ f } } )
{ \mathcal{ F } }_{ f f, \sigma }^{ \dagger}( { \bf k }, \omega_{ n } )
& = &
- V \sqrt{z} { \mathcal{ F } }^{ \dagger }_{ c f, \sigma }( { \bf k }, \omega_{n } )
      \nonumber \\
& &+
z^{2} { \Delta }^{ * }_{ k }( { \bf k } )
{ \mathcal{G } }^{ f f }_{\sigma }( { -\bf k }, \omega_{ n } )  \, ,
\nonumber \\
\label{SlaveFff}
\\
( i \omega_{n} - \tilde{ E_{ f } } ) 
{ \mathcal{G } }^{ f f }_{\sigma}( { -\bf k }, \omega_{ n } ) & = & V \sqrt{z} 
{ \mathcal{ G } }^{ c f }_{\sigma }( -{ \bf k }, \omega_{ n} )
\nonumber \\
& & + z^{2} \Delta_{k}
 { \mathcal{ F} }_{ f f, \sigma }^{ \dagger}( { \bf k }, \omega_{ n } )
\nonumber \\
& &+ 1 \, ,
\label{SlaveGff}
\end{eqnarray}
where the superconducting gap was already defined
in Eq. \ref{EqDelta}. Also note that
the above system of equations with the mapping
$ D_{ \sigma } = 1 $,
$ V \leftrightarrow \sqrt{ z } V $ 
and $ W \leftrightarrow z^{2} W $
is identical to the X-boson's result. 

\bibliography{apssamp}

\end{document}